\documentclass[a4paper,11pt]{article}
\pdfoutput=1
\usepackage{jheppub}

\usepackage[utf8]{inputenc}

\usepackage{amsmath}
\usepackage{amsfonts}
\usepackage{amssymb}
\usepackage{graphicx}
\usepackage{rotating}
\usepackage{pdflscape}
\usepackage{xspace}
\usepackage{xcolor}
\usepackage{mciteplus}
\usepackage{upgreek}
\usepackage{hyperref}
\usepackage[all]{hypcap}
\usepackage{ifthen}
\usepackage{authblk}
\usepackage{makecell}

\usepackage{lineno}

\newboolean{uprightparticles}
\setboolean{uprightparticles}{false} 

\def\DsTT{D_{s0}^*(2317)^+}
\def\DsTO{D_s^{*}(2112)^+}
\def\DsFE{D_{s1}(2460)^+}
\def\DsTS{D_{s1}(2536)^+}

\usepackage{xspace} 
\usepackage{upgreek}

\def\MagUp {\mbox{\em Mag\kern -0.05em Up}\xspace}

\ifthenelse{\boolean{uprightparticles}}%
{

 \def\Ppi         {\ensuremath{\uppi}\xspace}

 \def\PDelta      {\ensuremath{\Delta}\xspace}                 
 \def\PXi         {\ensuremath{\Xi}\xspace}                 
 \def\PLambda     {\ensuremath{\Lambda}\xspace}                 
 \def\PSigma      {\ensuremath{\Sigma}\xspace}                 
 \def\POmega      {\ensuremath{\Omega}\xspace}                 
 \def\PUpsilon    {\ensuremath{\Upsilon}\xspace}

 \def\PB      {\ensuremath{\mathrm{B}}\xspace}                 
                  
 \def\PD      {\ensuremath{\mathrm{D}}\xspace}

 \def\PK      {\ensuremath{\mathrm{K}}\xspace}

 \def\Pi      {\ensuremath{\mathrm{i}}\xspace}

 \def\Ps      {\ensuremath{\mathrm{s}}\xspace}

 \def\thebaroffset{0.0em}
}
{

 \def\Ppi         {\ensuremath{\pi}\xspace}

 \mathchardef\PDelta="7101
 \mathchardef\PXi="7104
 \mathchardef\PLambda="7103
 \mathchardef\PSigma="7106
 \mathchardef\POmega="710A
 \mathchardef\PUpsilon="7107
                  
 \def\PB      {\ensuremath{B}\xspace}                 
                  
 \def\PD      {\ensuremath{D}\xspace}

 \def\PK      {\ensuremath{K}\xspace}

 \def\Pi      {\ensuremath{i}\xspace}

 \def\Ps      {\ensuremath{s}\xspace}

 \def\thebaroffset{0.18em}
}
\newcommand{\offsetoverline}[2][\thebaroffset]{\kern #1\overline{\kern -#1 #2}}%

\makeatletter
\ifcase \@ptsize \relax
  \newcommand{\miniscule}{\@setfontsize\miniscule{4}{5}}
\or
  \newcommand{\miniscule}{\@setfontsize\miniscule{5}{6}}
\or
  \newcommand{\miniscule}{\@setfontsize\miniscule{5}{6}}
\fi
\makeatother

\DeclareRobustCommand{\optbar}[1]{\shortstack{{\miniscule (\rule[.5ex]{1.25em}{.18mm})}
  \\ [-.7ex] $#1$}}

\def\squark    {{\ensuremath{\Ps}}\xspace}

\def\pion   {{\ensuremath{\Ppi}}\xspace}
\def\piz    {{\ensuremath{\pion^0}}\xspace}
\def\pip    {{\ensuremath{\pion^+}}\xspace}
\def\pim    {{\ensuremath{\pion^-}}\xspace}

\def\KorKbar {\kern \thebaroffset\optbar{\kern -\thebaroffset \PK}{}\xspace}

\def\D       {{\ensuremath{\PD}}\xspace}

\def\DorDbar {\kern \thebaroffset\optbar{\kern -\thebaroffset \PD}\xspace}

\def\Dp      {{\ensuremath{\D^+}}\xspace}
\def\Dm      {{\ensuremath{\D^-}}\xspace}

\def\DpDm    {\ensuremath{\Dp {\kern -0.16em \Dm}}\xspace}

\def\Ds      {{\ensuremath{\D^+_\squark}}\xspace}
\def\Dsp     {{\ensuremath{\D^+_\squark}}\xspace}

\def\B       {{\ensuremath{\PB}}\xspace}

\def\BorBbar {\kern \thebaroffset\optbar{\kern -\thebaroffset \PB}\xspace}

\def\Bd      {{\ensuremath{\B^0}}\xspace}

\def\BdorBdbar {\kern \thebaroffset\optbar{\kern -\thebaroffset \Bd}\xspace}

\def\Bs      {{\ensuremath{\B^0_\squark}}\xspace}

\def\BsorBsbar {\kern \thebaroffset\optbar{\kern -\thebaroffset \Bs}\xspace}

\def\Y#1S{\ensuremath{\PUpsilon{(#1S)}}\xspace}

\def\LorLbar     {\kern \thebaroffset\optbar{\kern -\thebaroffset \PLambda}\xspace}

\def\to                 {\ensuremath{\rightarrow}\xspace}

\def\AT#1     {\ensuremath{A_{\mathrm{T}}^{#1}}\xspace}           

\def\C#1      {\ensuremath{\mathcal{C}_{#1}}\xspace}                       
\def\Cp#1     {\ensuremath{\mathcal{C}_{#1}^{'}}\xspace}                    
\def\Ceff#1   {\ensuremath{\mathcal{C}_{#1}^{\mathrm{(eff)}}}\xspace}        
\def\Cpeff#1  {\ensuremath{\mathcal{C}_{#1}^{'\mathrm{(eff)}}}\xspace}       
\def\Ope#1    {\ensuremath{\mathcal{O}_{#1}}\xspace}                       
\def\Opep#1   {\ensuremath{\mathcal{O}_{#1}^{'}}\xspace}                    

\newcommand{\aunit}[1]{\ensuremath{\text{\,#1}}}       

\newcommand{\tev}{\aunit{Te\kern -0.1em V}\xspace}
\newcommand{\gev}{\aunit{Ge\kern -0.1em V}\xspace}
\newcommand{\mev}{\aunit{Me\kern -0.1em V}\xspace}
\newcommand{\kev}{\aunit{ke\kern -0.1em V}\xspace}
\newcommand{\ev}{\aunit{e\kern -0.1em V}\xspace}
 
\newcommand{\mevc}{\ensuremath{\aunit{Me\kern -0.1em V\!/}c}\xspace}
\newcommand{\gevc}{\ensuremath{\aunit{Ge\kern -0.1em V\!/}c}\xspace}
\newcommand{\mevcc}{\ensuremath{\aunit{Me\kern -0.1em V\!/}c^2}\xspace}
\newcommand{\gevcc}{\ensuremath{\aunit{Ge\kern -0.1em V\!/}c^2}\xspace}

\def\gsim{{~\raise.15em\hbox{$>$}\kern-.85em
          \lower.35em\hbox{$\sim$}~}\xspace}
\def\lsim{{~\raise.15em\hbox{$<$}\kern-.85em
          \lower.35em\hbox{$\sim$}~}\xspace}

\def\tell1  {TELL1\xspace}
\def\ukl1   {UKL1\xspace}

\title{\boldmath Why do we not see the radiative decays of $D_{s1}^+(2536)$?}

\author[a,b]{Alex Bondar}

\affiliation[a]{Budker Institute of Nuclear Physics of SB RAS, 630090 Novosibirsk, Russia}
\affiliation[b]{Novosibirsk State University, 630090 Novosibirsk, Russia}

\emailAdd{A.E.Bondar@inp.nsk.su}

\abstract{This article is an attempt to extract information about the radiative width of the $\DsTS \to \Dsp \gamma$ decay from the results available in the literature and to discuss the consequences of such an attempt.}

\begin{document}

\maketitle
\flushbottom




\section{Introduction}

Since the experimental discovery of narrow P-wave states of the charm-strange quarkonia $D_{s0}(2317)^+$~\cite{BaBar:2003oey} and $D_{s1}(2460)^+$~\cite{CLEO:2003ggt}, discussions continue on the quark structure of these states~\cite{Cahn:2003cw,Barnes:2003dj,Kolomeitsev:2003ac,Chen:2004dy,Guo:2006fu,Guo:2006rp,Wang:2006uba,Bardeen:2003kt,Nowak:2003ra,Fu:2021wde,Tang:2023yls}. At the same time, much less attention is paid to the partners of these states $\DsTS$ and $D_{s2}(2573)^+$, which is probably due to the general consensus that these states are conventional quarkonia. At present, the discussions in literature are mainly over the mechanisms and magnitude of mixing in the physical state $\DsTS$ of ideal P-wave states with j=1/2 and j=3/2~\cite{Wu:2011yb}. As for the predictions of radiative widths for these states, they date, with rare exceptions, to the time period before 2003, when narrow P-wave states were just discovered. These predictions are usually based on the constituent quark model. Despite the presence of theoretical predictions of the radiative widths of $\DsTS$ and $D_{s2}(2573)^+$, there are no experimental data on these decays in literature~\footnote{ The statement in paper~\cite{Asratian:1987rb} about the observation of $\DsTS$  in the $\bar{\nu} N$ interaction in the $D_s^* \gamma$ decay channel is not credible, since the absence of signal in $D^* K$ is stated in the same paper.}. Perhaps this is due to the fact that the expected probabilities of radiative decays of these states are rather small, but there are not even upper limits on these probabilities. The absence of attempts to observe the radiative decays of $D_{s1}(2536)^+$ and $D_{s2}(2573)^+$ in the experiment seems to be an unfortunate omission. In this article, based on the results available in literature, an attempt is made to obtain an upper limit on the width of the radiative decay $\DsTS\to\Dsp\gamma$ and compare the obtained limit with the currently available theoretical predictions of the radiative widths of this decay.\\

\section{What can we get from experiment?}

As mentioned above, there are currently no direct upper limits on the radiative decays of $D_{s1}(2536)^+$. However, there is a detailed study by Babar~\cite{BaBar:2006eep} in which the products of the $D_{s0}(2317)^+$ and $D_{s1}(2460)^+$ production cross sections in $e^+e^-$ annihilation inclusive events and the decay probabilities of these states into a number of final states are measured. List of the measured cross sections is presented in Table~\ref{tb:yields}. 

\begin{table}[!h]
\caption{\label{tb:yields}Summary of the decay yield results~\cite{BaBar:2006eep}.
All cross sections are calculated for center-of-mass momentum $p^* > 3.2$~\gevc. 
The first quoted uncertainty for the central value
is statistical and the second is systematic. The limits
correspond to 95\% CL.}
\renewcommand{\baselinestretch}{1.3}
\begin{tabular}{llr@{$\:\pm\:$}r@{$\:\pm\:$}rr}
\multicolumn{2}{l}{Decay Mode} 
     & \multicolumn{3}{c}{Central Value (fb)} & Limit (fb)  \\
\hline
\multicolumn{6}{l}{$\sigma(\DsFE)\mathcal{B}(\DsFE\to X) \mathcal{B}( \Ds\to\phi\pip )$} \\
& $\Ds\piz$          & $ -1.0$ & $  1.4$ & $  0.1$ & $<  1.7$ \\
& $\Ds\gamma$        & $ 14.4$ & $  1.0$ & $  1.4$ &    ---   \\
& $\DsTO\piz$        & $ 41.6$ & $  5.1$ & $  5.0$ &    ---   \\
& $\DsTT\gamma$      & $  1.1$ & $  5.1$ & $  5.0$ & $< 15.2$ \\
& $\Ds\pi^0\pi^0$    & $  5.5$ & $  5.4$ & $  2.4$ & $< 28.5$ \\
& $\Ds\gamma\gamma$  & $  3.5$ & $  4.3$ & $  1.7$ & $< 13.2$ \\
& $\DsTO\gamma$      & $ -0.9$ & $  3.5$ & $  4.1$ & $<  9.7$ \\
& $\Ds\pi^+\pi^-$    & $  3.3$ & $  0.5$ & $  0.3$ &    ---   \\
\hline
\multicolumn{6}{l}{$\sigma(\DsTS)\mathcal{B}(\DsTS\to X) \mathcal{B}( \Ds\to\phi\pip )$} \\
& $\Ds\pip\pim$      & $  5.2$ & $  0.7$ & $  0.4$ & ---\\ 
\end{tabular}
\end{table}

\begin{figure}[!h]
\vskip -0.15in
\includegraphics[width=\linewidth]{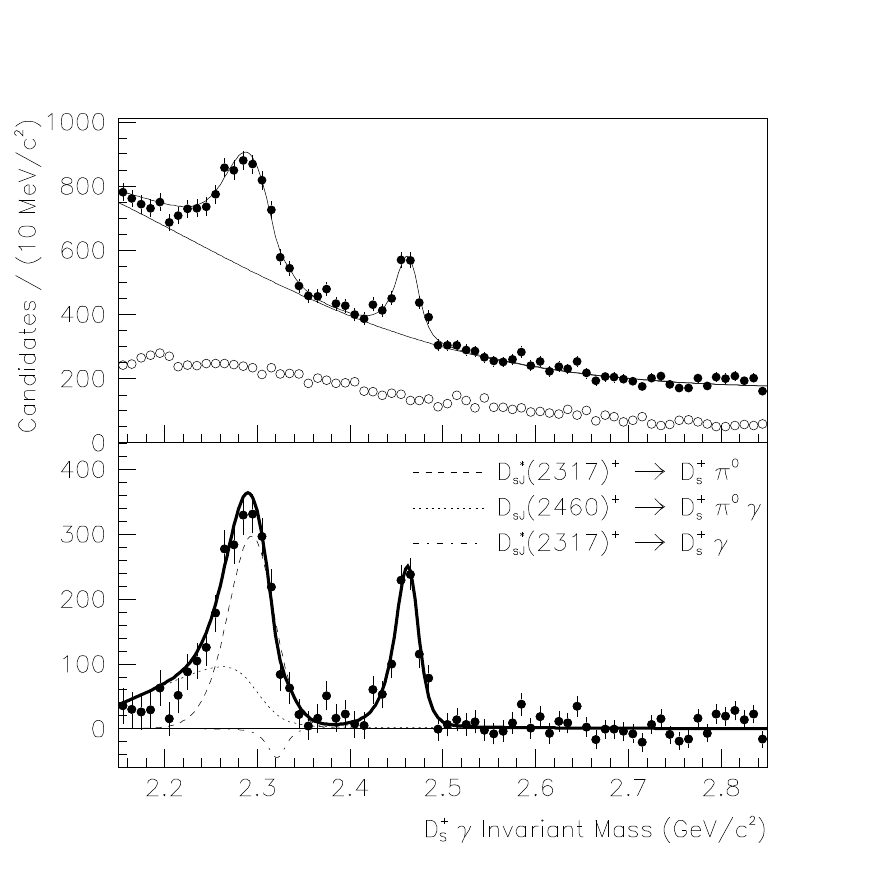}
\vskip -0.15in
\caption{\label{fg:dsgam.plotfitdatapaper}An example likelihood fit
to the $\Ds\gamma$ invariant mass distribution. The solid points in the top
plot are the mass distribution. The open points are the $\Ds$ sidebands,
scaled appropriately. The bottom plot shows the same data 
after subtracting the background curve from the fit. Various contributions
to the likelihood fit are also shown. The plot is taken from~\cite{BaBar:2006eep}.
}
\end{figure}

In Fig.~\ref{fg:dsgam.plotfitdatapaper} the distribution of experimental events over the $D_s \gamma$ invariant mass is shown, and the $\DsFE$ signal is clearly visible. The estimated raw $\DsFE$ yield from the fit is $920 \pm 60$ events. At the same time, there are no signs of signal at the $\DsTS$ mass. Assuming that the detection efficiency of the final state $D_s \gamma$ for $\DsFE$ and $\DsTS$ is the same, it is possible to conservatively estimate the number of detected $\DsTS$ decay events as less than 120 at the 95$\%$ confidence level (CL). Thus, using Table~\ref{tb:yields} results, we get the upper limit for the product of the $\DsTS$ production cross section and the decay probabilities $\DsTS \to D_s \gamma$, $D_s \to \phi \pi^+$, and $\phi \to K^+K^-$ as 1.9~fb. In this case, we do not take into account the contribution of the systematic error, since, according to Babar~\cite{BaBar:2006eep}  , the systematic error in determining the yield of $\DsFE \to D_s^+ \gamma$ events is multiplicative and amounts to 10$\%$. Using the known decay probability $D_s \to \phi \pi^+, \phi \to K^+K^-$ ~\cite{ParticleDataGroup:2022pth} equal to $(2.21 \pm 0.06)\%$, we find  the upper limit for the product of the inclusive $\DsTS$ production cross section and the $\DsTS\to D_s^+\gamma$ decay probability equal to 85~fb (the cross sections obtained by Babar~\cite{BaBar:2006eep} are given for reconstructed meson momenta $p^* > 3.2$~\gevc).

Now we use the fact that the inclusive cross section of $\DsTS$ production in $e^+e^-$ annihilation on $\Upsilon(4S)$ was measured by CLEO~\cite{CLEO:1993nxj} with following decays to $D^{*0} K^+$ and $D^{*+} K^0$. The cross section times branching ratio turns out to be 

$\sigma(e^+e^- \to \DsTS X)\times Br(\DsTS \to D^{*0} K^+) = (6.5 \pm 1.1 \pm 1.0)$~pb, 

$\sigma(e^+e^- \to \DsTS X)\times Br(\DsTS \to D^{*+} K^0) = (5.8 \pm 1.0 \pm 0.9)$~ pb. 

Since $D^{*0} K^+$ and $D^{*+} K^0$ are the dominant decay channels of $\DsTS$, the sum of these cross sections determines the total inclusive cross section for $\DsTS$ production with good accuracy. 

In the same article, CLEO measured the $\DsTS$ fragmentation function, which turned out to be significantly more rigid than in case of $D_s^+$ and $D_s^{*+}$ ($\epsilon = 0.014_{-0.005}^{+ 0.01} \pm 0.003$, where $\epsilon$ is the parameter of the Peterson fragmentation function~\cite{Peterson:1982ak}). Based on the measured fragmentation function, one can obtain the value of the inclusive $\DsTS$ cross section for $p^* > 3.2$~\gevc as $(9520 \pm 2000)$~fb.  Since the total width of $\DsTS $ is known and equals $(0.92 \pm 0.05)$~ MeV~\cite{ParticleDataGroup:2022pth} we can estimate the  $\DsTS \to D_s^+ \gamma$ and $\DsTS \to D_s^+ \pi^+ \pi^-$ decay probabilities and decay widths: 

$Br(\DsTS \to D_s^+ \gamma) < 0.009$, $\Gamma(\DsTS \to D_s^+ \gamma) < 8$~keV, 

$Br(\DsTS \to D_s^+ \pi^+ \pi^-)=(0.025 \pm 0.0065)$, 

$\Gamma(\DsTS \to D_s^+ \pi^+ \pi^-) = (23 \pm 6)$~keV.

\section{What can follow from this consideration?}
It is interesting to compare the resulting constraint on $\Gamma(\DsTS \to D_s^+ \gamma) $ with theoretical predictions~(Table~\ref{tb:width.predictions}). 

\begin{table}[!h]
\caption{\label{tb:width.predictions} Theoretical predictions of the width of the radiative $\Gamma(\DsTS\to D_s^+\gamma)$ decay.
}
\begin{center}

\begin{tabular}{lrr}
       & \multicolumn{1}{c}{  } &\multicolumn{1}{c}{  }\\
Paper & \multicolumn{1}{c}{ }&\multicolumn{1}{c}{$\Gamma(\DsTS \to D_s^+ \gamma) $ (keV)}\\
\hline
S.Godfrey~\cite{Godfrey:2005ww} &    &  15   \\
 J.~L.~Goity and W.~Roberts~\cite{Goity:2000dk}        &    &  25.2-31.1   \\
F.~E.~Close and E.~S.~Swanson~\cite{Close:2005se}   &    &  7   \\
N.~Green, W.~W.~Repko and S.~F.~Radford~\cite{Green:2016occ}              &    &  61.2   \\
S.~F.~Radford, W.~W.~Repko and M.~J.~Saelim`\cite{Radford:2009bs}       &    &  54.5   \\
S.~F.~Chen, J.~Liu, H.~Q.~Zhou and D.~Y.~Chen~\cite{Chen:2020jku}  &    &  18.18-18.85   \\
T.~Matsuki, K.~Seo~\cite{Matsuki:2012qc}&       & 27\\
J.~G.~Korner, D.~Pirjol and K.~Schilcher~\cite{Korner:1992pz}               &    &  1.6 $\pm$ 2.3   \\

\hline
This paper                &    &  $<$8 ($95\%$ CL)   \\
\end{tabular}

\end{center}
\end{table}

All the theoretical predictions for the $\Gamma(\DsTS \to D_s^+ \gamma) $ decay width shown in this table, except for two~\cite{Close:2005se, Korner:1992pz}, significantly exceed the obtained upper limit. If $\DsTS $ is indeed a quarkonium, then in the quark model approximation the width of the discussed decay is described by the well-known formula (see for example~\cite{Godfrey:2005ww}):
\begin{equation}
\Gamma(i \to f + \gamma) = \frac{4}{27} \alpha <e_Q>^2 \omega ^3 (2J_f + 1) |< 
 {^{2s+1}S_{J_f}} |r|\space  ^{2s+1}P_{J_i} >|^2 S_{if} \;,  
\label{eq:E1width}
\end{equation}
where $S_{if}$ is a statistical factor with $S_{if} = 1$ for the transitions between spin-triplet states and $S_{if} = 3$ for the transition between spin-singlet states, $<e_Q>$ is an effective quark charge given by
\begin{equation}
<e_Q> = \frac{m_qe_c - m_ce_{\bar{q}}}{  m_c + m_q } \;,
\label{eq:eQ}
\end{equation}
where $e_c = +2/3$ is the charge of the c-quark and $e_{\bar{q}} = +1/3$ is the charge of the s antiquark given in units of $|e|$, $m_c$ and $m_q$ are the masses of the c and s quarks, $\omega$ is the photon's energy, $\alpha$ is the fine-structure constant. The specific values of the c and s quark masses are fixed in the potential models from the description of the position of the quarkonium ground states. The matrix element $<S|r|P>$ is evaluated using the wave-functions calculated with the model potentials which describe the quark-antiquark interactions. The quark model in various modifications usually predicts well the widths of radiative transitions in quarkonium, so the calculation of the matrix element in case of $\DsTS$ can be considered quite reliable. However, as is known, in case of electric and magnetic dipole transitions of $(c\bar{s})$ states, the contributions of $c$ and $\bar{s}$ quarks to the transition amplitude significantly cancel each other. This can be seen from the formula~\ref{eq:eQ} for the effective electric charge $<e_Q>$. For example, for the ratio of quark masses $m_s/m_c = \frac{1}{2}$ the value of $<e_Q>$ is set to zero. The fact that in potential quark models the probability of $D_s^{*+} \to D_s^+ \gamma$ radiative transition can be overestimated by several times was noted earlier~\cite{Bardeen:2003kt}.

Assuming the quarkonium nature of $ \DsFE $ within the same model the upper limit can be estimated also for the radiative decay width $\Gamma(\DsFE \to D_s^+ \gamma) $, taking into account that the electric dipole transition also preserves the spin of the initial state:

\begin{equation}
\frac{\Gamma(\DsFE \to D_s^+ \gamma)}{\Gamma(\DsTS \to D_s^+ \gamma)} = \frac{\sin^2\theta}{\cos^2\theta} \left[\frac{\omega(\DsFE)}{\omega(\DsTS)}\right]^3 = 0.34\;,
\label{eq:radwidthratio}
\end{equation}
which corresponds to $ \Gamma(\DsFE \to D_s^+ \gamma) < 2.7$~keV~\footnote{However, it cannot be completely ruled out that the degree of the reduction in the E1 radiative transitions may be different for the states $P_{j=3/2}$ and $P_{j=1/ 2}$ due to unaccounted $1/m_c$ corrections.}, where $\theta$ is the mixing angle of the spin singlet and spin triplet initial state wave functions (in the heavy quark limit ${\sin^2\theta}/{\cos^2\theta} = 1/2$). Since we know the ratio of the widths $ \Gamma(\DsFE \to D_s^+ \pi^+ \pi^-)/\Gamma(\DsFE \to D_s^+ \gamma)=0.24 \pm 0.06$~\cite{ParticleDataGroup:2022pth}, we can also obtain a limit on the decay width $ \Gamma(\DsFE \to D_s^+ \pi^+ \pi^-) < 0.65$~keV, which looks quite strange in comparison with the previously obtained $\Gamma(\DsTS \to D_s^+ \pi^+ \pi^-)=23$~keV decay width. Apparently, there must be some special reasons why the decay widths in $ D_s^+ \pi^+ \pi^-$ of two pseudovector states of the charm-strange quarkonium that are close in mass are so different, or the initial assumption that $ \DsFE $ is a pure quarkonium state was incorrect. It is possible that a significant admixture of the molecular state ($D^*K$) removes the suppression of the effective charge in the radiative transitions of $\DsFE$ and thereby decreases the ratio of the three-body and  radiative decay widths. Indeed, the theoretical predictions for the $ \Gamma(\DsFE \to D_s^+ \gamma)$ decay in the molecular model give a width value of about $20-40$~keV~\cite{Colangelo:2005hv,Close:2005se,Fu:2021wde} and for the $ \Gamma(\DsFE \to D_s^+ \pi^+ \pi^-) = 16^{+7}_{-5}$~keV~\cite{Tang:2023yls}.


\section{Conclusion}

Based on the experimental data available in literature, we obtain an estimate of the partial decay width $\Gamma(\DsTS \to D_s^+ \pi^+ \pi^-) = (23 \pm 6)$~keV and an upper limit on the radiative decay width $ \Gamma(\DsTS \to D_s^+ \gamma) < 8$~keV (95\% CL). Within the framework of the constituent quark model, assuming the purely quarkonium nature of $\DsFE$, the corresponding widths ($ \Gamma(\DsFE \to D_s^+ \gamma) < 2.7$~keV and $\Gamma(\DsFE \to D_s^+ \pi^+ \pi^-) < 0.65$~keV) are estimated for this state. Based on this, it is concluded that $\DsFE$ state probably has a significant contribution from the molecular component $(D^{*}K)$, which is consistent with  earlier assumption made in the papers~\cite{Barnes:2003dj,Kolomeitsev:2003ac,Chen:2004dy,Guo:2006fu,Guo:2006rp} to explain the considerable difference of measured $\DsFE$ mass from the predicted value obtained within the potential models.

To confirm the conclusions made in this paper, it seems necessary to measure $\DsFE$ and $\DsTS$ mesons radiative decay probabilities and their inclusive production cross sections with a new level of statistical accuracy, which is already available in the Belle II  experiment, and to study in detail mechanism of these mesons' decays to $ D_s^+ \pi^+ \pi^- $.

\section{Acknowledgments}
I am grateful to Alexandr Milstein, Anton Poluektov, Arkady Vainshtein and Anna Vinokurova for useful discussions and valuable suggestions.

\bibliographystyle{JHEP}
\bibliography{note}

\end{document}